# Goals for teacher learning about energy degradation and usefulness

Abigail R. Daane, Stamatis Vokos, and Rachel E. Scherr

*Department of Physics, Seattle Pacific University, Seattle, Washington 98119, USA*

The Next Generation Science Standards (NGSS) require teachers to understand aspects of energy degradation and the second law of thermodynamics, including energy's availability and usefulness, changes in energy concentration, and the tendency of energy to spread uniformly. In an effort to develop learning goals that support teachers in building robust understandings of energy from their existing knowledge, we studied teachers' impromptu conversations about these topics during professional development courses about energy. Many of these teachers' ideas appear to align with statements from the NGSS, including the intuition that energy can be present but inaccessible, that energy can change in its usefulness as it transforms within a system, and that energy can lose its usefulness as it disperses, often ending up as thermal energy. Some teachers' ideas about energy degradation go beyond what is articulated in the NGSS, including the idea that thermal energy can be useful in some situations and the idea that energy's usefulness depends on the objects included in a scenario. Based on these observations, we introduce learning goals for energy degradation and the second law of thermodynamics that (1) represent a sophisticated physics understanding of these concepts, (2) originate in ideas that teachers already use, and (3) align with the NGSS.

## I. INTRODUCTION

The Next Generation Science Standards (NGSS) require teachers to understand several aspects of energy degradation and the second law of thermodynamics, including energy's availability and usefulness, changes in energy concentration, and the tendency of energy to spread uniformly [1,2]. Secondary teachers, for example, must be prepared to teach that "although energy cannot be destroyed, it can be converted to less useful forms," and that "uncontrolled systems always evolve toward more even energy distribution" [2]. These requirements are effectively new: although earlier national documents included similar statements [3,4], those documents were not as widely endorsed as the NGSS. The dual emphasis in the NGSS on both energy conservation (the first law of thermodynamics) and energy degradation (the second law of thermodynamics) can support teachers in integrating these two aspects of the energy concept, which are too often unconnected in physics instruction.

Learning goals for teachers should not only stem from the content requirements of the NGSS, but should also originate in the productive ideas that teachers already use. As researchers who aspire to support teachers in standards-ready teaching and learning about energy, we study teacher conversations about energy during energy-related professional development courses to find out what they already know, what they need to know, and in what ways their understanding is incomplete. We find that teachers in professional development courses spontaneously consider ideas related to energy's availability and degradation, without prompting from the course objectives or support from the instructors. For example, some teachers view energy as losing value during certain processes, even as they explicitly recognize that the total amount of energy is constant. Others articulate that the quality, usefulness, or availability of the energy may decrease when the energy changes form (for example, from kinetic to thermal) or when the energy disperses in space. None of the teachers we observe demonstrate a complete understanding of energy degradation and the second law of thermodynamics. Yet in many cases, teachers' impromptu discussions indicate a valuable conceptual basis for a model of energy that includes both conservation and degradation.

In what follows, we first review the physics of energy degradation and the second law of thermodynamics, including the treatment of these topics in the NGSS and how they relate to sociopolitical aspects of energy. We describe previous research on learners' ideas about energy degradation (Sec. III), explain the methodology used to gather and interpret the data and describe the context and background for the observations of K-12 teachers as learners (Sec. IV). By analyzing teachers' spontaneous discussions about energy usefulness, degradation, dispersion, and availability, we identify alignments between physics and teachers' intuitive ideas (Sec. V). Finally, we share learning goals that originate with teachers' productive ideas about energy degradation and the second law of thermodynamics and align with the new NGSS requirements (Sec. VI). These learning goals support teachers integrating disciplinary and everyday understandings of energy.

## II. THE PHYSICS OF ENERGY DEGRADATION

In everyday language, people may refer to certain quantities of energy as lost or used up, prompting concerns as to whether or not they are committed to the concept of energy conservation. Another possible interpretation is that people are referring to ideas related to energy degradation. Energy degradation depends on a specific system (which comprises all relevant objects in a specific scenario), a specified time evolution of that system, and a specified or putative final state. We use the term *scenario* to refer to an "energy story" involving the objects comprising the system that has a predetermined time development (e.g., a basketball rolls to a stop, or an incandescent bulb glows steadily). *Degraded* energy at time $t$, associated with the system of all relevant objects evolving from some initial state to a specified final state is energy at time $t$ that will not be available for the performance of work during the remaining time evolution of the system [5,6]. In order to avoid requiring our teachers to integrate models of force and energy prematurely, we define degraded energy equivalently as energy unavailable for the process of mechanical energy transfer (with the provisos of the previous sentence). For example, in a gasoline-powered car, thermal energy that dissipates to the environment as the engine runs is lost in that it cannot be used to propel the car; it is degraded. Energy change associated with the performance of work (or mechanical transfer) is related to *free* energy change in physics ("free" in the sense that it is available for use) [7]. The total energy is, at every instant, the sum of the degraded energy and the free energy (of the system, for a given scenario). Concerns about conserving energy (in the sociopolitical sense of guarding against energy waste) may be interpreted as concerns about preserving free energy.

In physics, energy degradation is associated with movement toward equilibrium in a quantity whose gradient can be harnessed for the performance of work (such as temperature, pressure, or concentration). When a partition is removed between a vacuum and a cube of gas, the gas expands from the area of high concentration into the volume that was a vacuum. This expansion process reduces the pressure difference between the two volumes and degrades the energy that was associated with the filled cube. The expansion also spreads energy more equitably through the system [9]. Energy can also spread through mixing: for example, when a hot gas and a cold gas come into contact with each other, the initial temperature gradient between them is reduced. In this case, the energy spreads in phase space by increasing the range of possible momenta of the particles. In real, irreversible processes, energy spreads within objects, to other objects, through space, by mixing, in phase space, or a combination of these [10]. This spreading is accompanied by an increase in entropy [9]. In other words, energy spreading, energy degradation, reduction of gradients, and entropy production are all features of real, irreversible processes. This co-occurrence prompts a degradation-oriented statement of the second law of thermodynamics: Energy degrades in irreversible processes.

Since energy degradation is defined relative to a specific set of objects interacting over a time interval through specified processes and with specified initial and final states (i.e., a specific scenario), changes in any of those parameters can change the status of the energy (from degraded to free, or vice versa). For example, thermal energy that accumulates in a car as a result of the engine running may be identified as degraded in the system consisting only of the car and the warm surrounding air, because it cannot be used to propel the car. However, that same thermal energy may be identified as free in a system that includes freezing surrounding air—a system that includes a distinct temperature gradient that could be used, in theory, to power some other process. The addition of new objects into the scenario, or reconsideration of the boundaries of the system, can introduce new gradients and increase the free energy of the system. Thus, degraded and free are not properties of units of energy; rather, they are qualities of the distribution of energy among interacting objects.

The NGSS and its parent document, the *Framework for K-12 Science Education* [2], reflect pressing societal concerns with energy topics in that energy is both one of the seven "cross-cutting concepts" that bridge disciplinary boundaries and one of the four "disciplinary core ideas" in the physical sciences. The primary learning goals about energy reflected in the NGSS are that energy is conserved, that it manifests in multiple ways, and that it is continually transferred from one object to another and transformed among its various forms. Within the topic of energy, the NGSS and the *Framework* particularly support concepts related to energy degradation and the second law of thermodynamics in their emphasis on energy-efficient solutions to societal problems. Energy degradation and the second law are treated in the sections titled "Conservation of energy and energy transfer" (PS3.B, where PS stands for Physical Science) and "Energy in chemical processes and everyday life" (PS3.D). Specific statements from the Framework regarding energy degradation include the following:

- "Energy in concentrated form is useful for generating electricity, moving or heating objects, and producing light, whereas diffuse energy in the environment is not readily captured for practical use." (PS3.D)
- "A system does not destroy energy when carrying out any process. However, the process cannot occur without energy being available. The energy is also not destroyed by the end of the process. Most often some or all of it has been transferred to heat the surrounding environment; in the same sense that paper is not destroyed when it is written on, it still exists but is not readily available for further use." (PS3.D)
- "The expression 'produce energy' typically refers to the conversion of stored energy into a desired form for practical use… When machines or animals 'use' energy (e.g., to move around), most often the energy is transferred to heat the surrounding environment." (PS3.D, endpoint for 5th grade)
- "Although energy cannot be destroyed, it can be converted to less useful forms–for example, to thermal energy in the surrounding environment." (PS3D, endpoint for 12th grade)
- "Uncontrolled systems always evolve toward more stable states—that is, toward more uniform energy distribution (e.g., water flows downhill, objects hotter than their surrounding environment cool down). Any object or system that can degrade with no added energy is unstable." (PS3B, endpoint for 12th grade)

The vision of the NGSS is for all students to gain "sufficient knowledge of the practices, crosscutting concepts, and core ideas of science and engineering to engage in public discussions on science-related issues, to be critical consumers of scientific information related to their everyday lives, and to continue to learn about science throughout their lives" [2]. The topic of energy degradation offers a particularly valuable opportunity to realize this vision. Energy conservation is central both in a sociopolitical sense and in the formal study of physics, but the term has a different meaning in each context. In physics, energy conservation refers to the idea that the same total quantity of energy is always present in any isolated system; energy is neither created nor destroyed. In the public consciousness, however, energy conservation refers to the idea that we have to guard against energy being wasted or used up; the energy available to serve human purposes is both created (in power plants) and destroyed (in processes that render it unavailable to us). Both of these ideas correspond to laws of thermodynamics: the first law (energy is conserved) and the second law (energy degrades). The NGSS can support teachers in integrating these two aspects of the energy concept, which are too often unconnected in physics instruction. The NGSS can also support teachers in recognizing the value of student ideas about energy degradation, ideas that have often been treated as misconceptions (see Sec. III).

## III. PREVIOUS PHYSICS EDUCATION RESEARCH ON ENERGY DEGRADATION AND THE SECOND LAW OF THERMODYNAMICS

### A. Research on teachers' ideas about energy degradation

Research regarding K-12 teachers' knowledge of energy in general is sparse. When the British education system moved to a National Curriculum, the teaching requirements for primary teachers increased to include energy concepts previously reserved for secondary education; in this context hour-long, in-depth interviews of 20 primary teachers revealed that many teachers described energy as able to be created and destroyed during certain processes [11]. For example, "one teacher felt that energy could be created and destroyed but that …'it depends what [situation] you're talking about'" [11]. Instead of identifying potential connections to free energy and degraded energy, researchers have usually identified these ideas as barriers to learning about energy. In Portugal, 16 secondary science teachers were observed in their classrooms while teaching a unit on heat, temperature, and energy. Researchers found that teachers used a mixture of both everyday language and formal descriptions of energy concepts to help students apply the information to the real world: for example, when discussing the energy crisis in class, one teacher stated, "there is always some energy which is not recuperated for new use," and another said, "when we run out of energy we have to eat" [12]. The idea that "there is always some energy which is not recuperated" can be a productive resource for learning about energy degradation. However, the researchers feared that this mixing of descriptions may have led to more confusion for the students [12]. A more recent study followed 20 secondary science teachers from Spain through a professional development course on energy degradation and into their own classrooms; these secondary teachers were found to use "inappropriate conceptual meanings" [13] of energy degradation including the following: (a) energy transfer or transformation indicates degradation, (b) degradation reduces total energy or occurs only when energy is not conserved, (c) internal energy is unrelated to degradation, and (d) degradation is heat, which teachers seemed to define as "a process of losing energy or losing the availability of energy" [13].

### B. Research on students' ideas about energy degradation

Although there is little research on K-12 teachers' understanding of energy, a number of investigations report on secondary and undergraduate students' ideas about energy. Introductory university students' conceptual understanding of physics is documented to be similar in some cases to K-12 teachers' understanding (see, e.g., [14,15]). It is reasonable to expect that this similarity would hold for energy degradation, a topic typically covered in upper-division physics courses; only secondary teachers specializing in physics might be expected to have taken such courses, and even among that subpopulation of K-12 teachers a physics major is rare [16]. Thus, prior research about undergraduate students' and even secondary students' ideas may inform our understanding of the variety of general ideas people have regarding energy degradation and the second law of thermodynamics.

Students' ideas about energy often include sociopolitical associations with energy sources (e.g., fuels and food) and consumers (e.g., cars and humans) that are not consistent with the ideas taught in physics instruction [17–27]. For example, one preliminary study highlights a student's description of energy as "something that can do something for us…say like gas or something" [20]. As in research on the ideas of teachers, such ideas are usually identified as barriers to learning about energy. Some researchers argue that these "obstinately persistent" ideas must be confronted to improve students' understanding of energy [24,26].

Students tend to apply their everyday ideas about energy (e.g., energy as fuel) to real world situations in lieu of energy topics from canonical physics [22,23,28–34]. Studies of secondary students in several countries, including Germany, the Philippines, and the United Kingdom, have elicited student responses to questions regarding the principle of energy conservation. One British study found that after instruction, 30% of students surveyed still referred to conservation as saving energy or recycling; over 50% are reported to have "considerable difficulties with the basic concept of energy and its related ideas, and their application to everyday situations" [30]. A German study reporting on 34 clinical interviews with 15–16 year old students who had completed four years of traditional physics courses found that a majority describe energy as a substance that is used up [33]. Though this idea can prompt concerns about energy conservation, it may be productive if it is interpreted as applying to free energy.

### C. Instructional claims based on prior research

Many researchers tout an increased focus on energy degradation in K-12 classrooms as a way to increase student understanding [22,24,26,31,33–38]. In one study, lessons that focused on energy degradation significantly increased student learning of the principle of energy conservation [26]. Some researchers recommend middle school curricula that include degradation [36,37], but there is little research into the effectiveness of this approach. Other researchers recommend a stronger emphasis on free energy in physics curricula [8,39–41]. The use of free energy as it relates to gradients in a quantity (e.g., temperature or pressure) can support causal reasoning about energy [39]. One study promotes the use of the term "fuel energy" for free energy, responding to secondary students' ideas regarding sources and users [41].

Emphasis on the second law of thermodynamics may also improve student understanding of energy in secondary education. Studies suggest the use of "energy degrades" as a K-12 appropriate conceptual version of the second law of thermodynamics [30,34,35,41,42]. For example, teachers can increase student understanding of energy conservation by also using the "Running Down Principle": "In all energy changes there is a running down towards sameness in which some of the energy becomes useless" [35].

## IV. RESEARCH CONTEXT

In this section we share our methods for data collection, instructional context, and theoretical framework.

### A. Data collection and episode selection

Our data include examples of teachers discussing ideas related to energy degradation and the second law of thermodynamics. The examples are from video records of professional development courses for K-12 teachers offered through Seattle Pacific University as part of the Energy Project, a five-year, NSF-funded project to develop and study teacher practices of formative assessment in the context of energy teaching and learning. These energy-centered professional development courses are documented with video, field notes, and artifact collection (including photographs of whiteboards, written assessments, and teacher reflections). In each course, teachers as learners are grouped into 4–8 small groups, and two groups are recorded daily. As researcher-videographers document a particular course, they take real-time field notes in a cloud-based collaborative document, flagging moments of particular interest and noting questions that arise for them in the moment. Later, the researcher-videographers or other members of the Energy Project identify video episodes to share with a research team. We use the term "episode" to refer to a video-recorded stretch of interaction that coheres in some manner that is meaningful to the participants [43]. These episodes are the basis for collaborative analysis, development of research themes, literature searches, and the generation of small or large research projects.

For this analysis, video episodes were identified through (1) initial observations by videographers and (2) a search for key terms in the field notes which could relate to energy degradation (e.g., entropy, spreading, diffusion, thermal energy, wasted). Selected episodes were watched several times to support the creation of detailed narratives and transcripts. On the basis of multiple viewings of the video episodes and analysis of the transcripts and narratives, we identified the productive ideas related to energy degradation and the second law of thermodynamics that appear in learners' spontaneous discussions. Twelve episodes were isolated and captioned to illustrate learner engagement with issues of energy degradation. These episodes are described in Sec. V.

### B. Instructional context

Our research is motivated by our experience, as instructors, that attention to learners' productive ideas is among the most powerful tools for facilitating growth. We find that in practice, attending to learners' ideas requires active engagement by both instructors and peers and stimulates learners' own resources for problem solving [44–47]. In courses offered by the Energy Project, instructors place a high priority on attending to learners' productive intuitions.

They listen to the disciplinary substance of learners' ideas, adapting and discovering instructional objectives in response to learner thinking [48]. As a result, each course has a unique trajectory that emerges from the interaction of learners' agency with instructors' judgment of what is worth pursuing [45,49].

The episodes identified for this study do not show instructors responding to the substance of teachers' ideas about energy degradation because at the time that these courses took place, instructors were attending to learners' ideas about the primary learning goals of the course: (1) Learners should be able to conserve energy locally in space and time as they track the transfers and transformations of energy within, into, or out of systems of interest in real-world scenarios, and (2) learners should be able to theorize mechanisms for energy transfers and transformations [50,51]. By studying video records of learners' conversations, we have come to recognize value in previously overlooked learner ideas. Learning goals that reflect these ideas will shape instructor attention in future courses. Video observations and measurements of effectiveness feed an iterative cycle of physics model development, improvement of instructional activities, and advancement of learning theory.

Energy Project professional development courses are centered on a learning activity called Energy Theater, an embodied representation based on a substance metaphor for energy [50–52]. In Energy Theater, each participant identifies as a unit of energy that has one and only one form at any given time. Groups of learners work together to represent the energy transfers and transformations in a specific physical scenario (e.g., a refrigerator cooling food or a light bulb burning steadily). Participants choose which forms of energy and which objects in the scenario will be represented. Objects in the scenario correspond to regions on the floor, indicated by circles of rope. As energy moves and changes form in the scenario, participants move to different locations on the floor and change their represented form. The rules of Energy Theater are as follows:

- Each person is a unit of energy in the scenario.
- Regions on the floor correspond to objects in the scenario.
- Each person has one form of energy at a time.
- Each person indicates their form of energy in some way, often with a hand sign.
- People move from one region to another as energy is transferred, and change hand sign as energy changes form.
- The number of people in a region or making a particular hand sign corresponds to the quantity of energy in a certain object or of a particular form, respectively.

Energy is associated with each object based on perceptible indicators that specify the state of that object. A snapshot of Energy Theater illustrates the energy located in each object at the instant of the snapshot, consistent with understanding energy as a state function.

### C. Theoretical framework

Both our professional development and our research take as a premise that learners have stores of intuitions about the physical world, informed by personal experience, cultural participation, schooling, and other knowledge-building activities [53–59]. Some of these intuitions are "productive," meaning that they align at least in part with disciplinary norms in the sciences, as judged by disciplinary experts [49,60,61]. Learners may only apply these intuitions episodically: at some moments of conversation with instructors and peers there may be evidence of productive ideas, whereas at other moments productive ideas may not be visible [62,63].

We conceptualize learning as a process of growth through which the "seeds" of learners' early ideas mature, through experience, to become more logical, coherent, consistent with observed evidence, and otherwise more fully scientific. Effective instruction, in this view, is instruction that provides favorable conditions for growth. This general conceptualization is common to many specific theories about teaching and learning [59,64–68]. Some research contrasts this general conceptualization with other conceptualizations that see learners as hindered by ideas that are fundamentally flawed, and instruction as repairing or replacing learners' ideas [61,69,70].

Overall, the literature on student learning of the first and second law of thermodynamics characterizes student ideas as different from and in competition with what they learn in science courses. Some argue that "students are evidently inadequately prepared and/or are unable to use the energy concept and the principle of the conservation of energy in order to explain simple experiments. They prefer to use explanations with which they are familiar from their environment" [29]. Even when concepts such as degradation are introduced into the classroom, researchers report that students lack the ability to connect their ideas to concepts covered in physics instruction, saying, for example, "It is obvious…that students are far from the physicist's conception of energy degradation" [33].

We share these researchers' sense of the importance of student ideas for instruction. However, we see these ideas not as obstacles to learning but as potentially productive resources for sophisticated understanding [55,58]. Research using a resources theoretical perspective has been conducted for the concept of energy [49,60,71], but not energy degradation. We argue that students' and teachers' everyday ideas about saving and wasting energy contain the seeds of correct canonical physics concepts now included in the NGSS: The intuition that energy can be "lost to us" is a productive idea as applied to irreversible processes in the real world. The outcome of our analysis is to propose learning goals for future professional development that originate from teachers' productive ideas and form a coherent concept of energy degradation. In the next section, we will share teachers' ideas that are productively aligned with aspects of energy degradation, and describe how those ideas can be combined to create a coherent understanding of the concept of energy degradation used in the NGSS.

## V. TEACHERS' IDEAS RELATED TO ENERGY DEGRADATION

Teachers in our courses have discussed ideas about energy that our instruction was not designed to support. We provide examples of teachers considering elements of energy degradation without explicit instruction or encouragement. Their ideas include the following: (A) energy can be present but inaccessible; (B) energy can lose its usefulness as it transforms within a system; (C) energy can lose its usefulness as it disperses; (D) energy tends to end as thermal energy; and (E) energy's usefulness depends on the choice of objects involved in the scenario (Fig. 1). Below we describe these ideas and give examples of their use by teachers as learners in our courses.

### A. Energy can be present but inaccessible

Some teachers describe energy being used up during a process, even as they explicitly acknowledge that the total amount of energy is constant. In the following episode, teachers discuss the energy involved when wind creates waves on water. Four elementary teachers (whose pseudonyms are Joel, Rosie, Hannah, and Marissa) decide that energy transfers from the wind to the water, and then try to determine what happens to the energy after that. Hannah states that the energy in the water waves is "not absorbed, because it has to continue on to go someplace." Marissa asks what happens to the energy if the wave hits a wall. Joel suggests the wave "goes through this mass and hits every individual particle." He asks, "Does every single thing take a little bit of energy away until it eventually dies off?" Joel's word "it" might refer to either the energy or the wave dying off. In the ensuing exchange, Rosie interprets Joel's question as a suggestion that the energy dies off. Rosie and Hannah then agree that the energy is "gone," but pause to clarify the meaning of that assertion.

- Energy can be present but inaccessible.
- Energy can lose its usefulness as it transforms within a system.
- Energy can lose its usefulness as it disperses.
- Energy tends to end as thermal energy.
- Energy's usefulness depends on the objects involved in the scenario.

FIG. 1. Teachers' ideas related to energy degradation.

Rosie: But it can't ever die though, right? Isn't that what we decided?
Hannah: No, but it can though, because look at batteries, a battery is stored energy, and when it's gone, it's gone.
Rosie: It's *gone!*
Hannah: It's gone somewhere, but it's gone.
Rosie: It's not really gone. It's just not there.
Instructor: It's gone somewhere.
Hannah: Right, exactly.
Instructor: It's just not in the battery anymore.
Rosie: Right, oh, but we just talked about dissipate, so that's the same thing as saying gone away from us.

After Joel's question Rosie counters, "but it can't ever die," possibly referring to conservation of energy. Hannah responds with a reference to batteries, which are sources of energy that are said to "die" when all possible chemical energy has been transformed to electrical energy. Hannah describes the energy in batteries as eventually being "gone." Though Hannah does not initially specify whether she means gone out of existence or gone to another location, she later says, "It's gone somewhere, but it's gone," supporting the location interpretation. Rosie affirms that "it's not really gone; it's just not there." She relates this idea to dissipation, which she seems to understand as a condition in which the energy is inaccessible ("gone away from us").

Rosie, Hannah, Joel, and Marissa retain their commitment to energy conservation when they assert that the energy of the water wave must "go somewhere" when it hits the wall. However, they also attempt to reconcile this commitment with their sense that the energy "goes away from us" as part of that process. In other episodes below, teachers recognize that energy may become inaccessible even as its quantity is unchanged (e.g., Dennis in Sec. V B, Vicki in Sec. V C, and Jean in Sec. V D).

### B. Energy can lose its usefulness as it transforms within a system

Teachers describe the usefulness and availability of energy during various energy processes. They distinguish between more or less useful energy and also explain how the usefulness changes during a process (e.g., energy becomes less useful when it transforms from kinetic to thermal energy). These informal descriptions of usefulness seem to correspond to the physics concept of degradation.

#### 1. Thermal energy is less useful

In our courses, some teachers describe a transformation into thermal energy as a loss of useful energy. This idea is also present in the NGSS, in which one standard (PS3.D) pronounces thermal energy a "less useful form" of energy. However, thermal energy can be useful in a situation where a temperature gradient drives a process (e.g., steam runs a turbine).

In the following episode, Dennis, a secondary teacher, distinguishes between useful energy and energy that has lost its usefulness. He also identifies the change in usefulness during a process. Responding to a question about a block sliding across the floor, Dennis says, "The molecules are heating up in the lower energy state. Somehow the system is going from a higher energy state to a lower energy state." The instructor shares a concern that Dennis's statement might violate the principle of energy conservation.

Instructor: I'm not sure exactly what you mean by "lower," given that my total energy has not changed.
Dennis: Oh! The usefulness of the energy is changing. It's going from, oh I can't think of the name, thermodynamic equilibrium?
Tom: Are you talking about entropy? You can't get the thermal back. It's gone.
Dennis: Yah, yah, the energy's there, but it's in terms of random motion. So, its quality, its usefulness is getting lost in the exchange.
Joe: The entropy, wouldn't you have less that can be used?
Dennis: What's that?
Joe: You have less that's available.
Tom: From a higher state to a lower state of entropy
Dennis: You have less that's available to you.
Joe: That would be a main condition of entropy
Doug: Mmhmm. Second law of thermo
Dennis: The energy stays the same, yet we're losing the usefulness.

Dennis states that a block sliding across the floor causes molecules to heat up and increases random motion. He says that the energy's availability, usefulness, and quality decrease during this process. When the instructor asks about energy conservation, Dennis clarifies that "the energy stays the same, yet we're losing the usefulness." In this statement, he distinguishes between the total amount of energy (constant) and the value of energy (decreasing). Dennis associates a "lower state" of energy with energy being "less available to you" and "less useful." His idea that the energy state is lowered is not aligned with disciplinary norms in physics. However, it conveys his sense that the energy's status is lowered by the transformation from kinetic energy to thermal energy. In this transformation energy disperses in phase space [9], which corresponds to an increase in degraded energy.

The other three secondary teachers, Tom, Joe, and Doug, support and extend Dennis's line of thinking. Tom adds, "You can't get the thermal back," and Joe suggests that there is less of something that can be used. The concepts of entropy and the second law of thermodynamics are proposed as potential extensions to the description of energy usefulness. However, the conversation shifts focus after Dennis's last statement and these connections are not discussed further.

In another course, a group of secondary teachers including Jennifer and Marta, discuss an Energy Theater scenario of a hand lowering a ball at constant speed. Jennifer indicates that a transformation into "heat" [72] makes the energy less useful. She says, "You know what? Seriously, the only place it [energy] could be going is heat cause it's obviously not going anywhere useful." Marta also proposes, "Let's just lose one person to heat, or something." Jennifer later suggests, "How about we have some of the people who are going from GPE [gravitational potential energy] to kinetic go away as heat or go into the Earth?" The recommendation that some people (who are chunks of energy in Energy Theater) should be lost or should go away as heat does not indicate that the law of conservation is being violated because in Energy Theater, all participants must remain a part of the scenario; they cannot physically disappear. Jennifer and Marta are instead suggesting that the people should go somewhere, possibly into the surroundings or into the loop of rope representing the Earth, similar to Hannah and Rosie's suggestion that energy has "gone away." Energy loss implies that the energy has moved away from its previous location and/or has become unavailable for its previous use.

### 2. Sound energy is less useful

To a lesser extent, we also observe that some teachers consider sound energy to have limited usefulness. Brice, an elementary teacher, tracks the energy of a ball that rolls to a stop and asks his peers, "If you could measure the sound coming off the ball, would that be a form of energy that's being lost, just like the heat energy?" When Brice asks if he can describe sound energy as lost, just like heat energy, he implies that the transformation into either thermal energy or sound energy renders that energy unavailable.

Similarly, a group of secondary teachers determine that when a falling object hits the ground the kinetic energy transforms into thermal energy and sound energy. The instructor probes further in the following short episode.

Instructor: Where does the sound go? I mean, so say some of it goes into
Roland: Well okay, air, vibrations, it will spread out into space
Ted: So less useful form
Leah: Right.

In this short excerpt, Ted appears to be contrasting sound energy to other energy forms, possibly the gravitational potential energy at the beginning of the scenario. It is not clear whether Ted refers to the energy as less useful because it is in the form of sound energy, because it spreads out into space, or for both reasons. Regardless, his sense that energy becomes less useful recalls the concept of energy degradation.

### C. Energy can lose its usefulness as it disperses

Several teachers describe a loss of energy usefulness being caused by energy dispersal into a larger physical space. These descriptions often come with a gesture indicating spreading. In what follows, an instructor asks a small group of elementary teachers where the energy goes after a falling object hits the ground. Marissa and Vicki share their contrasting ideas about energy conservation and energy dispersal in the subsequent conversation and gestures.

Instructor: Where does the sound energy and the kinetic energy in the air end up? We only took it to impact, but if we took it all the way, 'til the thing is down.
Vicki: A little, a little bit of heat [holds and moves the tips of her fingers in contact with each other]
Marissa: I get stuck because of that whole conversation that it doesn't, it never goes away.
Instructor: It doesn't go away.
Marissa: So we could keep following it and following it and following it, until it comes back around. [gesture; see Fig. 2]
Vicki: *Does* it come back around?
Marissa: Well ultimately,
Vicki: That's what I'm [shakes her head "no"]
Instructor: We said it doesn't go away. We didn't say it was necessarily cyclical. I guess that's kind of the question.
Marissa: Well I wouldn't say it comes back necess- the same energy comes back to the same object, but just thinking that with all of this energy just floating in space, it doesn't come or go. So it's all [gestures like she holds a basketball and shakes it]
Vicki: It keeps dispersing. I asked this question yesterday. Heat death of the universe. It juuuust keeeps goooing out. [gesture; see Fig. 3]
Instructor: So, how likely is it that I can somehow harness this thermal energy in the ground and do something with it?
Vicki: [shakes her head "no"]
Marissa: So you're saying take it to a place where people could use it?
Instructor: Or *not*.
Vicki: It gets broken up into smaller and smaller pieces, it's not gone, but it's not able to be use[inaudible]. I find it a little depressing. [laughter]

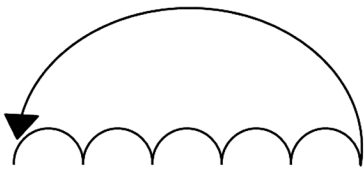

FIG. 2. The path of Marissa's hand movement for "So we could keep following it and following it and following it, 'til it comes back around."

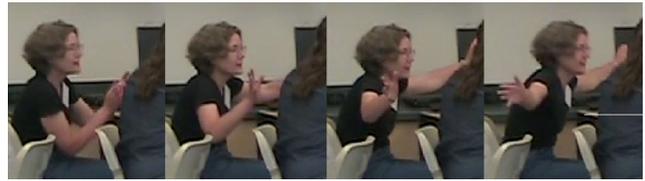

FIG. 3. Vicki's gesture for "It juuuust keeeps goooing out."

Marissa describes a cyclic model of energy conservation. She says that energy does not come or go, which implies that she is conserving energy. However, she shows a cyclic process of energy through the motion of her hand (Fig. 2) as she says, "So we could keep following it and following it and following it [hand makes small bounces to the right] 'til it comes back around [hand makes large arc back to the beginning position]." Marissa's description may not align with degradation because energy is reused for the same process in her model.

Vicki's model of the energy contrasts with Marissa's energy cycle. She describes the energy first as "a little bit of heat" and gestures by holding and moving the tips of her fingers in contact with each other. Then she uses a large gesture (Fig. 3) as she says, "It juuuust keeeps goooing out." During this gesture, she slows down each word and moves her hands slowly away from each other sideways, making a slight transverse wave motion up and down with her hands. After the gesture, the instructor asks if it is likely that this energy can be harnessed and Vicki shakes her head no.

In another elementary professional development course, Owen describes what happens to thermal energy that is produced by a hot plate. He says, "I think essentially it continues to travel [gesture; see Fig. 4]. The body of atmosphere that it's affecting becomes greater and greater and greater the more that it travels, so the impact of it becomes more and more negligible." During Owen's explanation, he pauses in silence to gesture outward with his hands (Fig. 4). His hands start close to his face with curled fingers and extend outward laterally while his fingers spread apart. He then states that the impact of the energy becomes more negligible as it spreads through a larger volume, similar to the idea that energy becomes less useful as it disperses.

Rosie, in the episode analyzed in Sec. VA., also discusses dissipation. She suggests that energy in a battery dissipates and equates that to energy going away. She states, "We just talked about dissipate, so that's the same thing as saying gone away from us." Rosie's gesture provides more information about what she means by "dissipate" (Fig. 5). Rosie's hands start together and move up and outward away from each other as she talks about the energy going away from us. Similar to Vicki and Owen, her hands spread out during this motion. Rosie's idea and her hand motions about dissipating energy align well with the idea of spreading energy to a less accessible location.

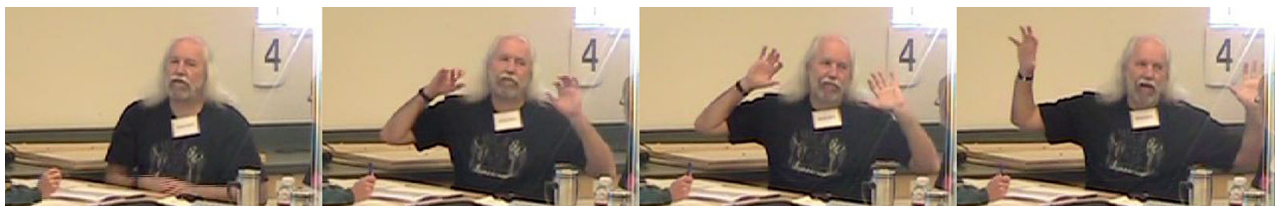

FIG. 4. Owen's gesture after he states, "I think it essentially continues to travel."

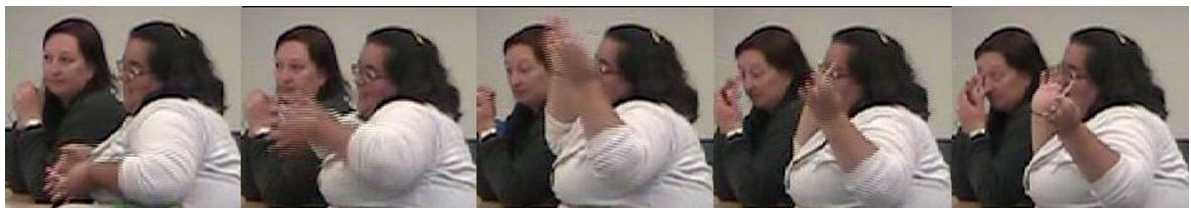

FIG. 5. Rosie's gesture for "We just talked about dissipate, so that's the same thing as saying gone away from us." (Rosie is holding a lollipop.)

This process of losing available energy through dissipation is closely aligned with the dispersal and degradation of energy.

### D. Energy tends to end as thermal energy

In previous sections, teachers describe thermal energy and sound energy as less useful forms of energy. In many scenarios teachers award thermal energy the lowest possible status of any energy form. Some teachers view a transformation into thermal energy as a terminal process, describing thermal energy as a "dead end" or the final form of energy. This idea aligns well with the second law of thermodynamics in that the energy in a given scenario experiences overall degradation. However, thermal energy does not have to be permanently degraded energy. In the following dialogue, a small group of elementary teachers considers where thermal energy goes when boiling water on a hot plate. Owen reminds the group (Jean, Karen, and Brianna) of a conclusion the class made previously that all energy ends in the form of thermal energy. A conversation in response to this idea about the terminal nature of thermal energy ensues.

Owen: Well, the idea that came up yesterday was that ultimately it all goes
Jean: to thermal!
Owen: [inaudible] to thermal.
Karen: All the energy ends there.
Brianna: It's just sitting there.
Jean: It's just sitting there! Yah, so how long does it sit there?[Short digression until Jean reiterates her description of the hot plate scenario upon the arrival of the instructor.]
Jean: I was saying that when I have done units on energy, so I have taken a hot plate, turned it on, and boiled water. And so we're talking about the energy transfers and transformations going on, but we always stop at heat energy and then we stop, like it's over.
Owen: When everything's warm.
Jean: Like the energy has, it's complete,
Instructor: [inaudible]
Jean: Yah, like it's done.[15 second discussion about how this idea has not come up in Jean's class]
Karen: Then you're sitting there wondering where
Jean: It's not really done because it's always there.
Karen: But weren't you asking at what point is it not, thermal energy anymore? At what point is it not there?
Jean: My question was: how long can energy stay the same thing in the thermal stage? Is it *days* or is that unrealistic? Or, I mean does it, because as soon as it changes, like with the wind, some of it is going to become sound energy and I don't know. It just made me wonder, what the life, the shelf-life is of thermal?
Owen: Yah, is it the shelf-life of Twinkies? Or
Jean: Yah! [Laughter]
Instructor: Forever.
Owen: And with yesterday's proposition that all energy ultimately is converted or transferred into thermal. So then, how long does thermal energy last?
Jean: Or if you were going to categorize energy would um, the largest percentage be thermal, and this percentage is sound, and this percentage is kinetic and this
Instructor: And that's energy in the entire universe?
Jean: Yah, yah, if you categorize it. It seems like it would mostly be thermal then, cause that's where everybody, everything ends.

In describing the result of a transformation into thermal energy, these teachers state that thermal energy "is just sitting there," and that once thermal energy is produced, this energy process is "complete," "over," or "done." They further generalize that "ultimately" all energy transforms into thermal energy and thermal energy is where "everything ends." Their idea that overall degradation occurred during the process is valid. These teachers also correctly view thermal energy in the air as unable to transfer or transform further in this isolated system (consisting of the hot plate, boiling water, and the air in the classroom). In physics, many common energy scenarios used when teaching the principle of energy conservation end in thermal energy (e.g., a sled sliding to a stop, a damped oscillator, a car braking). However, the idea that thermal energy is a permanently degraded form of energy could only be true if the entire Universe were in thermal equilibrium. If the system consisting of only the hot plate and the air in a classroom also included the cold surroundings of the outdoors (i.e., if a new gradient were introduced to the system), thermal energy in the air could transform and be used to perform work again.

In another course for elementary teachers, Brianna asks Brice and Bart about energy conservation and they conclude that all energy ends in the form of thermal energy.

Brianna: Somewhere in the back of my mind energy is never lost. Is that true? Energy is never, it just changes from one form to another and it never goes away. Is that true?
Brice: That is true.
Bart: I think that's true. [Laughter] It always changes into heat I think.
Brianna: Eventually?
Bart: It always degrades into heat. Yeah.

Bart first confirms that energy is conserved and then states that energy "always changes" and "degrades" into "heat." Bart's term heat might refer to thermal energy or radiation. Regardless, the use of the words "eventually" and "always" implies an overall degradation of the energy.

In a course for secondary teachers, Nancy, Ron, and Lucy track the energy of a hand pushing a ball vertically downwards under water. They conclude that energy cannot transform once it becomes thermal energy.

Instructor: Under what circumstances can I go from this form to that form? What must be going on in order for me to be doing that, and is that relevant?
Nancy: It's so interesting to me that intuitively I think in terms of things going from gravitational to thermal, or kinetic to thermal and not the other way around and that thermal is not going to be becoming something else.
Lucy: Kind of a dead end? That means things stay as thermal or?
Nancy: Yah? Maybe?

Nancy articulates that energy has a unidirectional tendency. She states that thermal energy "is not going to be becoming something else." Lucy suggests that thermal energy might be a dead end. Later in the conversation, Ron reiterates Nancy's idea, stating, "Because these T's (units of thermal energy) can't go back into G's (units of gravitational potential energy) can they?" Nancy agrees and suggests that the amount of thermal energy always increases, stating, "And of course there's my little idea that T's always get more." Although it is not the case that thermal energy cannot transform, nor is it true that thermal energy always increases, energy does degrade in spontaneous processes (the second law of thermodynamics), consistent with Nancy's intuition. Other teachers make similar statements. Toni, a secondary teacher discussing the Energy Theater for a light bulb asks, "Isn't heat our dead end?" [50]. Tom (Sec. V B) states, "You can't get the thermal back. It's gone."

Elementary and secondary teachers, both those who are new to our courses and those returning for subsequent professional development, suggest that transformation into thermal energy is a terminal process. This idea surfaces in a variety of energy scenarios: Owen and Jean discuss water boiling on a hot plate, Lucy and Nancy consider a hand pushing a ball into water, Brianna and Bart search for energy examples outside the classroom, Toni describes the energy in a light bulb, and Tom considers the energy scenario of a hand that pushes a block along the floor. The prevalence of this idea may be due to disciplinary emphasis on scenarios in which thermal energy is the endpoint of a process (e.g., friction). However, thermal energy is not always a dead end. The use of counterexamples in physics instruction might help highlight the fact that thermal energy can be used to perform work.

### E. Energy's usefulness depends on the objects involved

Above, Jean describes many scenarios, such as water boiling on a hot plate or a ball slowing to a stop, that end in thermal energy. She asks, "How long does [the thermal energy] sit there?" Her question implies that the thermal energy in the scenario does change eventually. Later, she brings up wind as a possible mechanism for transforming the thermal energy into sound energy. In questioning the finality of the situation, Jean touches on a subtle idea that energy degradation is not a property of each unit of energy but depends on the objects and processes in a particular scenario. For a given Energy Theater scenario, teachers tend to choose objects that constitute an isolated system (where all energy involved stays in that group of objects). Although many teachers in our courses distinguish between useful and less useful energy in the isolated system, few discuss how this designation depends on the choice of objects involved in the scenario.

In the following conversation, Jennifer, a secondary teacher, first reflects that Energy Theater and Energy Cubes, (another Energy Tracking Representation; see [51]), neglect to take into account energy usefulness and Irene, Kate, and Abdul respond to her ideas. The teachers then discuss the conditions under which thermal energy is useful and begin to compare scenarios involving different objects.

Jennifer: I was thinking about how in both Energy Theater and in the Energy Cubes that they had us use a different symbol or a different letter depending on what type of energy is represented. But the concept that you can't go backward, that once you have some of the energy transferred to the floor or the air as heat, that kids might think that you can, that that's equally reclaimable, or equally can be converted back to a more useful form of energy.

Abdul: Be reversible.

Jennifer: Reversible. So I wondered, what if you introduce the idea, or maybe kids could come up with the idea that in Energy Theater, the more useful the form of energy is, the taller you stand or something and every time the energy becomes less useful, like if it's sound or heat or something like that, that they shrink.

Irene: Especially as heat because it's wasted, part of entropy, that's why it's [gestures with hand brushing away]

Jennifer: Yeah.

Kate: Although heat can be useful, because if you burn coal then heat is useful, and that's what you're getting, right?

Jennifer: Right, but, what's useful, is, burning coal heats water, the steam turns the turbine, so what's really useful is the mechanical energy of the turbine moving.

Abdul: Amount of control could be useful, amount of control, when you can control this amount of heat could be useful. But in our case, you cannot control the heat energy.

Irene: Exactly.

Jennifer recognizes that Energy Theater and Energy Cubes do not address the "concept that you can't go backward." She states that "once you have some of the energy transferred to the floor or the air as heat that kids might think … that's equally reclaimable, or equally can be converted back to a more useful form of energy." To remedy this situation, she suggests that "in Energy Theater, the more useful the form of energy is, the taller you stand or something and every time the energy becomes less useful, like if it's sound or heat or something like that, that [the students] shrink." Although this addition of growing and shrinking could distinguish between free and degraded energy, we have not adopted this idea of shrinking into Energy Theater due to concerns about appearing not to conserve energy.

Jennifer's initial insights motivated our search for productive ideas about degradation. First, her description of the inability of energy to go backward, or convert back to a more useful form of energy, aligns well with the unidirectionality of irreversible processes and overall degradation of the energy in the system. Jennifer also distinguishes between more and less useful forms of energy, giving heat and sound as examples of less useful energy. These ideas parallel the formal physics distinction between free and degraded energy. Finally, Jennifer specifies that a change occurs in the usefulness of the energy when "the energy transferred to the floor or the air as heat." This loss of useful energy through the process of energy transfer describes energy degradation.

The discussion that follows Jennifer's observations suggests that teachers do not agree on whether or not thermal energy can be useful energy. First, Irene agrees with Jennifer that heat is not useful, saying heat is "wasted, part of entropy." Kate does not agree that thermal energy is always a less useful form. She counters, "Although heat can be useful, because if you burn coal then heat is useful, and that's what you're getting, right?" Jennifer's response to Kate indicates that she defines useful energy differently: she explains that coal produces useful "mechanical energy" in a turbine, aligning well with the formal physics definition of free energy. However, canonical physics would also identify the thermal energy from coal burning as useful because it transforms into mechanical energy. Finally, Abdul introduces yet another definition of useful, stating that thermal energy is useful "when you can control this amount of heat." Abdul implies that since, in this case, "you cannot control the thermal energy," it is not useful.

In this episode, teachers do not reach consensus about what useful energy is and what defines less useful or wasted energy. The lack of agreement highlights that whether energy is free or degraded depends on the scenario. If the scenario includes the coal, steam, and turbine, then thermal energy can be useful. However, if there is no turbine, then the thermal energy in the steam cannot be used for the performance of work. In other words, whether or not you can control or use the thermal energy depends on the chosen set of objects in a system. The disagreement between teachers about the status of thermal energy may have been because each had a different scenario in mind. Kate noted that it is possible to change the status by changing the scenario, and Abdul noted that the "amount of control" over the thermal energy determines its usefulness.

To summarize, Jennifer identifies the usefulness of energy, (i.e., one sociopolitical aspect of energy), as a concept that is missing in the representations. Teachers begin to address the idea that the usefulness of energy depends on the objects that are included in a given scenario. However, even for a small group of teachers, controversy exists regarding the definition of useful energy and in what (if any) situations thermal energy is considered useful. They touch on the idea that the status of energy as degraded or free is not a property of thermal energy; rather, the status of energy depends on the scenario (or the choice of objects involved in an isolated system), and the corresponding gradients that may drive energy transfer and transformation.

## VI. LEARNING GOALS FOR ENERGY-FOCUSED TEACHER PROFESSIONAL DEVELOPMENT

In the impromptu conversations described above, teachers view energy as losing value or becoming inaccessible during certain processes, while remaining constant in amount. In particular, teachers express that the quality, usefulness, or availability of energy may be reduced when it changes form or disperses in space. These observations, along with previous literature, are the basis for identifying teacher learning goals for energy degradation and the second law of thermodynamics (summarized in Fig. 6) that (1) represent a sophisticated physics understanding of these concepts, (2) originate in ideas that teachers already use, and (3) align with the NGSS.

### A. Teachers will distinguish between degraded energy and free energy in specific scenarios

Though teachers in our courses rarely use *degraded* or *free* to describe energy, they often describe energy as losing usefulness or accessibility. For example, Dennis (Sec. V B) describes thermal energy in his scenario as having less quality, usefulness, and availability than the initial kinetic energy and distinguishes between degraded and free energy; Ted (Sec. V B) states that sound energy is a less useful form of energy; and Jennifer (Sec. V E) suggests that students should indicate degradation by shrinking down in height. The distinction of energy as having more or less value is also made by Rosie, Vicki, Jean, and Owen (Secs. V A, V C, and V D, respectively). The NGSS distinguish degraded and free

---

- Teachers will distinguish between degraded energy and free energy in specific scenarios.
- Teacher will identify changes in the status of energy as they track the transfers and transformations of energy within isolated systems.
- Teachers will equate the total energy in a system at an instant to the sum of the degraded energy and the free energy.
- Teachers will identify the occurrence of overall energy degradation.
- Teachers will associate energy degradation with movement of a quantity towards equilibrium.
- Teachers will recognize that the identification of energy as degraded or free depends on the choice of the objects involved.

FIG. 6. Learning goals for energy degradation and the second law of thermodynamics.

---

energy in similar terms: "Energy in concentrated form is useful… whereas diffuse energy in the environment is not readily captured for practical use" (PS3.D). In sum, free ("useful") energy should be distinguished from degraded ("diffuse") energy that cannot be used to perform work in a specific set of objects [75].

### B. Teachers will identify changes in the status of energy as they track the energy transfers and transformations in isolated systems

Learners will not only distinguish between degraded energy and free energy, but also describe the circumstances associated with conversions from free to degraded energy and identify the mechanisms by which the changes take place. Energy degradation is often associated with movement towards equilibrium in some quantity, and can occur by means of dissipative processes such as friction. Several teachers in our courses identify energy degradation while analyzing energy dissipation. Dennis (Sec. V B) describes bulk movement transforming into random motion as a process by which energy becomes less useful (i.e., degraded). In Sec. V C, teachers describe the spreading of energy to a larger volume as the mechanism for a decrease in the accessibility or availability of energy. Many of the teachers in Sec. V D (e.g., Jean) describe transformation as a process that decreases the usefulness of energy. Finally, Jennifer identifies a decrease in usefulness as energy transforms into sound or thermal energy in Sec. V E. The NGSS address energy degradation processes as follows:

"A system does not destroy energy when carrying out any process. However, the process cannot occur without energy being available. The energy is also not destroyed at the end of the process. Most often some or all of it has been transferred to heat the surrounding environment; in the same sense that paper is not destroyed when it is written on, it still exists but is not readily available for further use." (PS3.D)

The NGSS use the idea of "availability" to describe the changing status of energy during a process (a transfer or transformation). If the energy is available (free), the process will occur; "without energy being available," however, "the process cannot occur."

### C. Teachers will equate the total energy in a system at an instant to the sum of the degraded energy and the free energy

This statement adds a key corollary to the principle of energy conservation: While the total energy of any isolated system remains the same, the relative amounts of degraded and free energy may change. Teachers in our episodes often explicitly state that energy is conserved, even as they are describing a loss of useful (i.e., free) energy. Rosie in Sec. V A points out that the total energy does not change in an isolated system, even as the energy becomes inaccessible (or has "gone away from us"). In Sec. V B, Dennis states that the energy is still there, even as its usefulness is getting lost. Vicki (Sec. V C) states that energy is not gone (indicating energy conservation) even as she laments that it becomes more spread out and less able to be harnessed. Finally, Jean (Sec. V D) argues that the thermal energy is "not really done, because it's always there," implying that although the energy seems to be no longer able to transfer or transform, it is still conserved. In the NGSS, this concept is addressed with the statement that "although energy cannot be destroyed, it can be converted into less useful forms" (PS3.D). This statement has the potential to reconcile the everyday meaning of conservation with the physics meaning, in that a decrease in free energy corresponds to sociopolitical energy being used up.

### D. Teachers will identify the occurrence of overall energy degradation

Though the total amount of energy in an isolated system is unchanged regardless of what physical processes may take place, the amount of free energy decreases during many physical processes. In other words, during many physical processes, energy degrades. This is a statement of the second law of thermodynamics that we see as particularly appropriate for K-12 teachers and students. Teachers in our courses recognize that energy becomes less useful as it goes through processes. In Sec. V D, Jean identifies the tendency for energy scenarios to end in thermal energy, which, in equilibrium, is degraded. The idea of energy reaching equilibrium is seen in statements describing thermal energy as a "dead end" or where energy "ends up." A similar idea is suggested in Sec. V C in Vicki's statement about the "heat death of the Universe." The NGSS statement of this concept is as follows:

"Any object or system that can degrade with no added energy is unstable. Eventually it will change or fall apart, although in some cases it may remain in the unstable state for a long time before decaying (e.g., long-lived radioactive isotopes)" (PS3.B).

This description aligns well with the idea of overall energy degradation. However, the NGSS use the term "degradation" to describe what happens to objects and systems, not energy.

Inherent in the previously identified statements of the second law of thermodynamics are the seeds of understanding entropy concepts. While explicitly teaching about entropy using the abstract mathematics that normally characterizes learning about entropy is not our goal at the secondary level, we are alert to opportunities to help teachers make valuable connections from energy to other concepts that can be constructed from everyday experience, including concepts associated with entropy. The discipline of physics stands to benefit from teachers' insightful conceptualizations of these concepts in terms that will be of use to them and their students.

### E. Teachers will associate energy degradation with movement of a quantity towards equilibrium

Free energy is associated with a difference (gradient) in some quantity potentially associated with work, and degraded energy with a lack of gradient in that quantity. Although teachers we observed did not yet identify the quantity whose gradient decreases and describe the corresponding energy degradation as that quantity moves towards equilibrium, this idea is foreshadowed in Sec. V C in the descriptions of energy dispersal. For example, Vicki says that as energy spreads and disperses, it is not likely to be harnessed, aligning with the idea that degraded energy cannot be used for work. Dennis (Sec. V B) refers to thermal equilibrium and Rosie (Sec. V A) mentions dissipation as the reason for a loss of accessible energy. The gestures that Vicki, Owen, and Rosie use during their explanations also illustrate dispersal. The NGSS address this concept by describing the tendency towards uniformity and stability:

"Uncontrolled systems always evolve toward more stable states-that is, toward more uniform energy distribution within the system or between the system and its environment (e.g., water flows downhill, objects that are hotter than their surrounding environment cool down)," (PS3.B).

While our learning goal associates energy degradation with the evolution of a quantity (e.g., pressure, concentration, temperature) towards equilibrium, the NGSS describe the evolution of a system towards a "more uniform energy distribution," implying a spreading in physical space. In many cases, these two conceptualizations are equivalent. We believe that associating energy degradation with the equilibration of a specific physical quantity helps to demonstrate how and why that energy distribution occurs.

### F. Teachers will recognize that the identification of energy as degraded or free depends on the choice of objects in the scenario

Teachers in Energy Project instruction discussed an aspect of energy degradation that goes beyond the material set out in the NGSS: the status of energy as useful or wasted depends on the objects in the scenario. Teachers in Sec. V E use this resource to highlight the relative nature of degradation when they discuss the meaning of useful energy and compare thermal energy in different scenarios. Jennifer and Kate describe how thermal energy may be useful in some situations and not in others. Implicit in these descriptions is the change of objects in the system. While the NGSS encourage learners to identify the objects that interact in a given energy scenario and track the energy as it transfers among those objects, as well as emphasizing the importance of designing energy-efficient systems, the standards do not require learners to understand that the status of energy as degraded or free depends on the choice of objects.

In addition, when Jennifer and Kate describe how thermal energy may be more or less useful in different scenarios (Sec. V E), they recognize that thermal energy can be useful in scenarios with a temperature gradient between the system of interest and the surrounding environment. The NGSS, in contrast, tend to implicitly identify thermal energy as not being useful:

- "[Energy] can be converted into less useful energy forms—for example, to thermal energy in the surrounding environment" (PS3.D).
- "Most often some or all of [the energy at the end of a process] has been transferred out of the system in unwanted ways (e.g., through friction, which eventually results in heat energy transfer to the surrounding environment)" (PS3.D).

In these statements, thermal energy is the only form described as "unwanted," "not readily available," and "less useful." The NGSS do not address the idea that thermal energy can be useful in the case of a temperature gradient between the surrounding environment and the system. Additionally, the NGSS do not describe how the introduction of a new object into a scenario can change the status of energy from degraded to free if the new object is not in equilibrium with the other objects. Degraded and free are not properties of units of energy; rather, they are qualities of the distribution of energy among interacting objects. These ideas can contribute to a coherent conceptual framework unifying everyday experiences and canonical physics.

## VII. CONCLUSION

The NGSS support a broad and comprehensive energy model including aspects of energy degradation and the second law of thermodynamics. Teachers are required to understand concepts including energy's availability and usefulness, changes in energy concentration, and the tendency of energy to spread uniformly. We identified episodes in our K-12 professional development courses in which teachers as learners show evidence of productive resources for understanding energy. Teachers spontaneously (without prompting from the course objectives or support from the instructors) considered not only the amount and forms of energy involved in physical processes, but also ideas related to the energy's availability and degradation that align with statements from the NGSS. Some teachers view energy as losing value during certain processes, even as they explicitly recognize that the total amount of energy is constant. Others articulate that the quality, usefulness, or availability of the energy may decrease when the energy changes form (for example, from kinetic to thermal) or when the energy disperses in space. Teachers also shared ideas not explicitly addressed in the NGSS about the usefulness of thermal energy and the context dependence of energy degradation. Although individual teachers in our courses had productive resources for learning about energy, they did not demonstrate a holistic understanding of the concept of energy degradation and the second law of thermodynamics. Based on these episodes and on previous literature, we developed learning goals that stem from teachers' existing conceptual resources.

Our aim is to support teachers in building a sophisticated understanding of energy in physics and society, one that is useful for them in everyday experiences as well as responsible to corresponding topics in formal physics and the NGSS. We expect the implementation of our learning goals in teacher professional development to increase the relevance of energy instruction in professional development, K-12 classrooms, and university contexts. This, in turn, may support awareness of the role of energy in human and natural systems, informing decisions to conserve, prepare, and make responsible energy choices [76].

## ACKNOWLEDGMENTS

We gratefully acknowledge all the elementary and secondary teachers who have participated in Energy Project courses for their generosity in making their own' reasoning accessible to the Energy Project team. We are grateful to Seattle Pacific University's Physics Education Research Group, including A. D. Robertson, S. B. McKagan, L. S. DeWater, L. Seeley, and K. Gray, and the University of Maryland Physics Education Research Group, especially E. F. Redish, B. D. Geller, and V. Sawtelle, for substantive discussions of this work. We also appreciate B. W. Harrer and V. J. Flood for their assistance with gesture analysis, and J. Haglund and S. Kanim for their assistance with the manuscript. This material is based upon work supported by the National Science Foundation under Grants No. 0822342 and No. 1222732.

[1] NGSS Lead States, *Next Generation Science Standards: For States, By States*, (The National Academies Press, Washington, DC, 2013).
[2] H. Quinn, H. Schweingruber, and T. Keller, *A Framework for K-12 Science Education: Practices, Crosscutting Concepts, and Core Ideas* (National Academies Press Washington, DC, 2012).
[3] American Association for the Advancement of Science and National Science Teachers Association, *Atlas of Science Literacy: Project 2061* (AAAS, Washington, DC, 2001).

[4] American Association for the Advancement of Science, *Benchmarks for Science Literacy 1993* (Oxford, New York, 1993).

[5] R. D. Knight, *Physics for Scientists and Engineers: A Strategic Approach: with Modern Physics* (Pearson, Addison-WesleyReading, MA, 2008).

[6] Degraded energy as defined here is therefore not a state function.

[7] Free energy as defined here is not a state function but rather the difference between state functions and may correspond to the work-related part of Gibbs or Helmholtz free energy changes depending on specific conditions. Our use of free energy corresponds to exergy [8], a term not widely used in physics instruction.

[8] H. T. Odum, *Environment, Power, and Society for the 21st Century: The Hierarchy of Energy* (Columbia University Press, New York, NY, 2007), p. 418.

[9] H. S. Leff, Thermodynamic entropy: The spreading and sharing of energy, Am. J. Phys. 64, 1261 (1996).

[10] Not all energy that spreads spatially is associated with irreversibility and energy degradation. For example, compressed springs that are arranged radially as spokes around a fixed center may be released, pushing blocks radially outward on a horizontal frictionless surface: the spatially localized energy in the springs spreads radially outward, but could bounce back from a fixed circular obstacle and recompress the springs.

[11] C. Kruger, Some primary teachers' ideas about energy, Phys. Educ. 25, 86 (1990).

[12] M. Louisa et al., Teachers' language and pupils' ideas in science lessons: Can teachers avoid reinforcing wrong ideas?, Intl. J. Sci. Educ. 11, 465 (1989).

[13] R. Pinto, D. Couso, and R. Gutierrez, *Using Research on Teachers' Transformations of Innovations to Inform Teacher Education: The Case of Energy Degradation* (Wiley Periodicals, Inc., New York, 2004), p. 38.

[14] L. C. McDermott, P. S. Shaffer, and C. Constantinou, Preparing teachers to teach physics and physical science by inquiry, Phys. Educ. 35, 411 (2000).

[15] A. D. Robertson and P. S. Shaffer, University student and K-12 teacher reasoning about the basic tenets of kinetic-molecular theory, Part I: Volume of an ideal gas, Am. J. Phys. 81, 303 (2013).

[16] Transforming the Preparation of Physics Teachers: A Call to Action. *A report by the Task Force on Teacher Education in Physics (T-TEP)*, edited by D. E. Meltzer, M. Plisch, and S. Vokos (APS, College Park, MD, 2012).

[17] J. Bliss and J. Ogborn, Children's choices of uses of energy, Eur. J. Sci. Educ. 7, 195 (1985).

[18] J. Solomon, Learning about energy: How pupils think in two domains, Eur. J. Sci. Educ. 5, 49 (1983).

[19] R. Trumper, Being constructive: An alternative approach to the teaching of the energy concept—part one, Intl. J. Sci. Educ. 12, 343 (1990).

[20] D. M. Watts, Some alternative views of energy, Phys. Educ. 18, 213 (1983).

[21] A. Brook, and R. Driver, *Aspects of Secondary Students' Understanding of Energy*, full report (University of Leeds: Centre for Studies in Science and Mathematics Education, Leeds, 1984).

[22] R. Duit, Learning the energy concept in school—Empirical results from The Philippines and West Germany, Phys. Educ. 19, 59 (1984).

[23] R. Duit, Energy conceptions held by students, and consequences for science teaching., in *Proceedings of the International Seminar "Misconceptions in Science, and Mathematics"*, edited by H. Helm and J. D. Novak (Ithaca, N.Y.: Cornell University, Ithaca, NY, 1983), pp. 316–321.

[24] J. Solomon, Messy, contradictory and obstinately persistent: A study of children's out-of-school ideas about energy, Sch. Sci. Rev. 65, 225 (1983).

[25] G. Nicholls and J. Ogborn, Dimensions of children's conceptions of energy, Intl. J. Sci. Educ. 15, 73 (1993).

[26] J. Solomon, Teaching the conservation of energy, Phys. Educ. 20, 165 (1985).

[27] R. Driver et al., *Making Sense of Secondary Science: Research into Children's Ideas* (Routledge, New York, NY, 1994).

[28] R. Driver and L. Warrington, Students' use of the principle of energy conservation in problem situations, Phys. Educ. 20, 171 (1985).

[29] R. Duit, Understanding energy as a conserved quantity - Remarks on the article by R. U. Sexl, Eur. J. Sci. Educ. 3, 291 (1981).

[30] H. Goldring and J. Osborne, Students' difficulties with energy and related concepts, Phys. Educ. 29, 26 (1994).

[31] P. Lijnse, Energy between the life-world of pupils and the world of physics, Sci. Educ. 74, 571 (1990).

[32] M. E. Loverude, C. H. Kautz, and P. R. L. Heron, Student understanding of the first law of thermodynamics: Relating work to the adiabatic compression of an ideal gas, Am. J. Phys. 70, 137 (2002).

[33] S. Kesidou and R. Duit, Students' conceptions of the second law of thermodynamics—an interpretive study, J. Res. Sci. Teach. 30, 85 (1993).

[34] J. Solomon, How children learn about energy or does the first law come first?, Sch. Sci. Rev. 63, 415 (1982).

[35] J. Solomon, *Getting to Know About Energy: In School and Society* (The Falmer Press, Bristol, PA, 1992).

[36] N. Papadouris and C. P. Constantinou, A philosophically informed teaching proposal on the topic of energy for students aged 11–14, Sci. & Educ. 20, 961 (2011).

[37] N. Papadouris, C. P. Constantinou, and T. Kyratsi, Students' use of the energy model to account for changes in physical systems, J. Res. Sci. Teach. 45, 444 (2008).

[38] J. Solomon, Learning and evaluation: A study of school children's views on the social uses of energy, Soc. Studies Sci. 15, 343 (1985).

[39] J. Ogborn, Energy, change, difference, and danger, Sch. Sci. Rev. 72, 81 (1990).

[40] J. Ogborn, Energy and fuel: The meaning of "the go of things", Sch. Sci. Rev. 68, 30 (1986).

[41] K. A. Ross, Matter scatter and energy anarchy; the second law of thermodynamics is simply common experience. Sch. Sci. Rev. 69, 438 (1988).

[42] R. Ben-Zvi, Non-science oriented students and the second law of thermodynamics, Intl. J. Sci. Educ. 21, 1251 (1999).

[43] B. Jordan and A. Henderson, Interaction analysis: Foundations and practice, J. Learn. Sci. 4, 39 (1995).

[44] D. L. Ball, With an eye on the mathematical horizon: Dilemmas of teaching elementary school mathematics, Elem. School J. 93, 373 (1993).

[45] D. Hammer, Discovery learning and discovery teaching, Cognit. Instr. 15, 485 (1997).

[46] C. R. Rogers, On Becoming a Person—A Psychotherapists View of Psychotherapy (Constable, London, 1961).

[47] S. B. Empson and V. J. Jacobs, Learning to listen to children's mathematics, in International Handbook of Mathematics Teacher Education, edited by P. Sullivan (Sense Publishers, Rotterdam, NL, 2008), pp. 257–281.

[48] J. E. Coffey, D. Hammer, D. M. Levin, and T. Grant, The missing disciplinary substance of formative assessment, J. Res. Sci. Teach. 48, 1109 (2011).

[49] D. Hammer, F. Goldberg, and S. Fargason, Responsive teaching and the beginnings of energy in a third grade classroom, Rev. Sci. Math. ICT Educ. 6, 51 (2012).

[50] R. E. Scherr, H. G. Close, E. W. Close, V. J. Flood, S. B. McKagan, A. D. Robertson, L. Seeley, M. C. Wittmann, and S. Vokos, Negotiating energy dynamics through embodied action in a materially structured environment, Phys. Rev. ST Phys. Educ. Res. 9, 020105 (2013).

[51] R. E. Scherr et al., Representing energy. II. Energy tracking representations, Phys. Rev. ST Phys. Educ. Res. 8, 020115 (2012).

[52] A. R. Daane, S. Vokos, and R. E. Scherr, Energy Theater, Phys. Teach. 52, 291 (2014).

[53] D. Hammer, Epistemological considerations in teaching introductory physics, Sci. Educ. 79, 393 (1995).

[54] A. A. diSessa, Toward an epistemology of physics, Cognit. Instr. 10, 105 (1993).

[55] E. Duckworth, The Having of Wonderful Ideas, and other essays on teaching and learning, (Teachers College Press, New York, NY, 1996).

[56] A. R. Daane, S. Vokos, and R. E. Scherr, Conserving energy in physics and society: Creating an integrated model of energy and the second law of thermodynamics, AIP Conf. Proc. 15, 114 (2013).

[57] A. Elby, Helping physics students learn how to learn, Am. J. Phys., Phys. Educ. Res. Suppl. 69, S54 (2001).

[58] D. Hammer, Student resources for learning introductory physics, Am. J. Phys. 68, S52 (2000).

[59] J. Dewey, Experience and Education (Simon and Schuster, New York, 1938).

[60] B. W. Harrer, V. J. Flood, and M. C. Wittmann, Productive resources in students' ideas about energy: An alternative analysis of Watts' original interview transcripts, Phys. Rev. ST Phys. Educ. Res. 9, 023101 (2013).

[61] D. Hammer, More than misconceptions: Multiple perspectives on student knowledge and reasoning, and an appropriate role for education research, Am. J. Phys. 64, 1316 (1996).

[62] T. G. Amin, A cognitive linguistics approach to the layperson's understanding of thermal phenomena, in Conceptual and Discourse Factors in Linguistic Structure, edited by A. Cenki, B. Luka, and M. Smith (CSLI Publications, Stanford, CA, 2001) p. 27.

[63] A. Gupta, D. Hammer, and E. F. Redish, The case for dynamic models of learners' ontologies in physics. J. Learn. Sci. 19, 285 (2010).

[64] J. J. Rousseau and B. Foxley, Emile: Or, On Education (Floating Press, Auckland, NZ, 2009).

[65] J. S. Bruner, The Process of Education (Harvard Univ. Press, Cambridge, 1960).

[66] M. Montessori, The Discovery of the Child (Ballantine Books, New York, 1978).

[67] J. Piaget, and B. Inhelder, The Psychology of the Child (Basic Books, New York, 1969).

[68] L. S. Vygotsky, Thought and Language (MIT Press, Cambridge, MA, 1986).

[69] J. P. Smith, A. A. diSessa, and J. Roschelle, Misconceptions reconceived: A constructivist analysis of knowledge in transition, J. Learn. Sci. 3, 115 (1994).

[70] D. Hammer, Misconceptions or p-prims: How may alternative perspectives of cognitive structure influence instructional perceptions and intentions, J. Learn. Sci. 5, 97 (1996).

[71] D. Hammer, and E. van Zee, Seeing the Science in Children's Thinking: Case Studies of Student Inquiry in Physical Science (Heinemann, Portsmouth, NH, 2006).

[72] Learners in our courses often use "heat" or "heat energy" to refer to a form of energy indicated by temperature (what we call thermal energy), rather than a transfer of energy driven by temperature difference (what we call heat) [50,73,74]. Unless otherwise mentioned, we interpret "heat" or "heat energy" as referring to thermal energy.

[73] P. A. Kraus and S. Vokos, The role of language in the teaching of energy: The case of "heat energy.", Washington State Teachers' Assoc. J (2011), http://www.spu.edu/depts/physics/documents/WSTA_KrausVokos.pdf.

[74] J. Warren, The teaching of the concept of heat, Phys. Educ. 7, 41 (1972).

[75] In a K-12 context, we do not expect teachers to distinguish among subtypes of free energy (e.g., Gibbs and Helmholtz).

[76] Science Education Resource Center, The principle for teaching energy awareness. 2013 April 14; http://cleanet.org/clean/literacy/energy.html.